# Influence of the photon beam position on incident SR intensity at SXMCD endstation [*]

GUO Yu-Xian(        )[1; 1)], XU Peng-Shou(        )[2]

[1] Department of Mathematics & Physics, Anhui University of Architecture, Hefei 230022, China;
[2] National Synchrotron Radiation Laboratory, University of Science & Technology of China, Hefei 230029, China

**Abstract:** A small fluctuation of the photon beam position will affect the characteristics of the synchrotron radiation (SR) intensity and polarization when it enters the end-station through the related beamline. In this paper, by changing the electron orbit equilibrium position in the vertical direction, we have measured the corresponding changes in the absorption strength of the SR with a gold mesh in different chopper aperture positions. It is found that for three aperture positions, the absorption intensity of the gold mesh shows a good Gaussian distribution as the photon beam position moves, while the ratio of the SR intensity passing through the upper and lower apertures shows a monotonous variation. This suggests a new method for estimating the circular polarization degree of SR originating from the bending magnet based on our current measurement.

**Key words:** polarization degree measurement, chopper, synchronous radiation (SR), X-ray magnetic circular dichroism (XMCD), X-ray absorption

**PACS:** 41.50.+h, 29.27.Hj, 29.20.dk

## 1  Introduction

X-ray magnetic circular dichroism (XMCD) refers to the dichroism phenomenon,

[*] Supported by the National Natural Science Foundation of China (21271007, 10274073) and Post-doctoral Research Start-up Funding of Anhui University of Architecture (K02553).
1) E-mail: guo_yuxian@163.com

which is generated for the different absorption of X-ray with the left-handed and right-handed circular polarization by a magnetized sample. Recently, with the rapid development of synchrotron radiation technology, it is possible to investigate the core energy level absorption edge of the 3d or 4f for different magnetic materials by using the XMCD method. Accordingly, this experimental technique has been widely applied in the fields of inorganic chemistry and bioinorganic chemistry. Compared with the traditional magnetic techniques, the XMCD technique shows obvious advantages: firstly, it has element-selected property; secondly, the related calculation is simple due to the transition of the core energy level. Thus combined with the XMCD experiment and the theoretical calculation, it is able to determine the spin and orbital magnetic moment of the specific atoms through proper approximation [1]. Based on the XMCD method, the magnetic anisotropy information of the magnetic sample and element-specific hysteresis loop in the complicated sample system [2] can be directly obtained. Furthermore, the clear magnetic microphotograph could be gained through the combination of the XMCD method and microtechnique [3], which is meaningful for the complicated sample system study from the microscopic magnetic moment aspect. It is known that the XMCD signal is closely associated with the external magnetic field as well as the circular polarization of the SR, thus it is very important to obtain the accurate circular polarization degree of SR during experiments. However, the SR from the bending magnet is always not stable and the SR polarization degree will be fluctuated, which will affect the XMCD measurement directly. In previous literature, people adopted many methods [4-6] to measure the polarization of SR. Aiming at Hefei Light Source (HLS), Sun [7], etc. designed a split photon beam position monitor to measure the light stability of the NSRL U18 Beamline. This apparatus can reflect the changes of beam position where detector is placed, but it gives little information about the beam direction as well as the accurate SR polarization degree. Obviously, a convenient and effective method for the SR polarization detection is desired and essential for the XMCD experiment.

In this paper, we have investigated the influences of photon beam position on the incident SR intensity, and attempted to develop an effective method at Hefei Light

Source (HLS) for the SR polarization monitor when the synchrotron beam position is fluctuated. By studying the relationship between the changes in height of electron orbit and the strength ratio of the emergent light passing through the upper and lower apertures, the SR polarization degree of the SXMCD endstation can be evaluated without delay.

## 2    Experiment

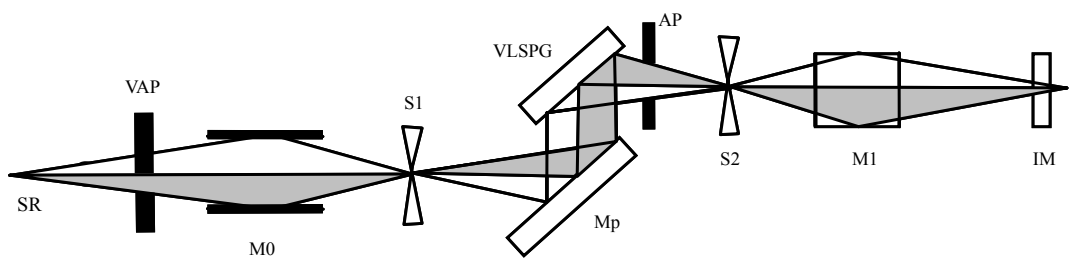

**Fig. 1.** Light path diagram of XMCD beamline, in which VAP refers to chopper, M0 and M1 refer to front-mounted focusing mirror and rear-mounted focusing mirror respectively; S1 and S2 refer to entrance slit and exit slit of the beamline respectively; plane mirror Mp and varied line-space plane grating VLSPG constitutes the monochromator of the beamline; IM refers to the light intensity monitoring room.

A standard XMCD beamline [8] normally includes chopper, front-mounted focusing mirror, entrance slit, monochromator, exit slit, refocusing mirror and light intensity monitor (Fig. 1). Monochromator is the key part of a beamline. For the XMCD Beamline in NSRL, the SR light is monochromized by a varied line-spacing plane grating monochromator (VLS-PGM) and refocused by a toroidal mirror, which has good anaberration function. The chopper is used to alternately select left-handed and right-handed circular polarized light at the upper and lower positions of the electron orbit. It is composed of chopping spin driving mechanism, function actuating element, light beam selection sliding mechanism, diaphragm differential mechanism, light strength monitoring device, cavity, observation window as well as precision regulation mechanism out of the four-degree of freedom cavity. The switching between the circular polarized light and the linear polarized light could be realized conveniently through switching the aperture position. The synchrotron light requires an online real-time monitor due to the characteristic of time and random drifting of the electron orbit. According to the real situation of the XMCD experiment, the gold mesh with 90% of transmissivity was used to monitor the incident light intensity in front of the sample. The monitoring of the incident light intensity was realized through measuring the sample current of the gold mesh. This signal manifests the relative strength of the incident light during the practical sample measurement process.

The beam position could be adjusted by changing the operating parameters with a convex rail system composed of three magnets. As the beam position moved 0.1-unit (equivalent to 75μm) each time, we measured the gold mesh absorption strength with two cycles at the upper, middle and lower positions of the aperture respectively.

## 3    Results and discussion

3.1    Changes in SR intensity caused by the movement of the photon beam position

Due to the switching time among different apertures, the signal collection at the gold mesh is not synchronous. Since the synchrotron light follows the decay characteristic with time, this out-synchronism will cause extra errors. To maintain this "synchronism", it is necessary to measure the absorption signal of the gold mesh at upper, middle, lower, middle and upper positions of the aperture respectively. So we can reduce the decay influence of the SR source by averaging the result of the two-cycle measurements.

As the electron orbit gradually moves up at a step of 0.1-unit, the gold mesh absorption strength at three different positions of the aperture is shown in Fig. 2. It could be found that the three curves show a Gaussian-like trend. The fitting process was carried out by Gaussian model and the $\chi^2/dof$ value was adopted to evaluate the fitting effect. Briefly, we define the light intensity passing through the upper, middle and lower apertures as $I_{up}$, $I_{middle}$ and $I_{down}$, respectively. According to the results of repeated measurements, the measurement errors, about ±0.3, were determined. Considering the degree of freedom (dof), n-3=17, it is easy to obtain the $\chi^2/dof$ values of Gaussian fitting as 0.94, 0.77 and 0.93 for $I_{up}$, $I_{middle}$, and $I_{down}$, respectively, which shows a perfect Gaussian distribution. By combining with the light path of the synchrotron light, we make the following analysis: for the synchrotron light intensity originating from the bending magnet is subject to the Gaussian distribution in the vertical direction, it also shall be subject to Gaussian distribution at the entrance slit in the vertical direction; as the photon beam position moved up gradually, the facula at the entrance slit moved down. For the position of entrance slit was fixed, the light, which coming from different parts of the facular in vertical direction, passed through the entrance slit successively, thus the light intensity passing through the slit should show up the changes similar to the Gaussian distribution; from Fig. 2, it could be found that the change trend of light intensity at the gold mesh is basically identical to that at the entrance slit theoretically. It indicates that the fluctuation of the photon beam position has little influence on the light transmission efficiency of the optical elements after entrance slit.

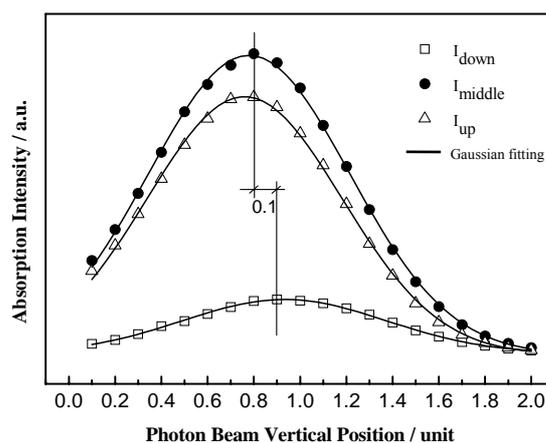

**Fig. 2.**  Changes in the absorption strength of gold mesh with the moving of the vertical position of photon beam for the different aperture positions. Herein, solid lines are Gaussian fitting results.

From Fig. 2, it can also be found that there is a big difference in the relative strength of the three curves; in which $I_{up}$ and $I_{middle}$ are strong while $I_{down}$ is weak. Combined with the analysis of the position of optical elements and the light path of synchrotron light, we infer that it results from the inconformity in height of the equilibrium position of electron orbit and the middle aperture of chopper; from Fig. 2, it could be obtained that the average photon beam position is located in height between the upper and middle apertures while it is far away from the lower aperture.

Furthermore, regardless of small difference, the corresponding peak values of these three curves in Fig. 2 are basically the same. As the SXMCD beamline adopts VLS-PGM, the total monochromator is highly capable in terms of anaberration. The three curves are obtained after the synchrotron light passes through all optical elements. The basic conformity of their peaks value occurring manifests the good anaberration function of the overall light beam system.

3.2 Relationship between the moving of the photon beam position and the SR polarization degree

From the measurement result in Fig. 2, the corresponding relationship between the photon beam position and the values of $I_{up}/I_{down}$ could be obtained easily, as shown in Fig. 3. As the photon beam position gradually moves up, the corresponding ratio would be reduced monotonically.

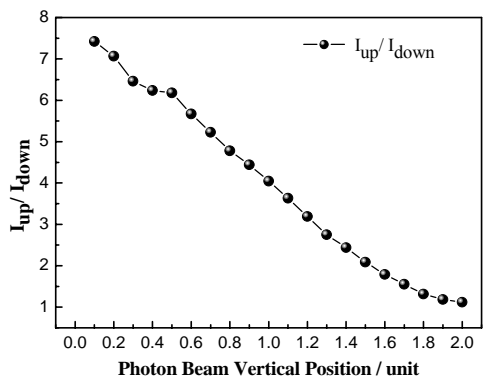

**Fig. 3.** The corresponding relationship between the light intensity ratio at upper and lower positions of the aperture and the vertical position of photon beam.

Furthermore, the data of chopper-up and chopper-middle were measured for two times during each two cycles of the experiment. It is found that the two times' coincidence effect is very good (not listed in Fig. 2); it indicates that, as the signals were obtained through a same aperture, the changes of the synchrotron light flux caused by the time decay could be basically ignored during the measurement process.

The synchrotron radiation is mainly polarized in the way that the electric vector parallels the acceleration vector. That is, the radiant electric vector sent out by the electron running in the circular orbit always points to the circular center. As far as the radiation from bending magnets is concerned, that eradiates along the electron orbit is the linearly polarized light while that diverges from the electron orbit is the elliptically polarized light. Even though the radiation from bending magnets could provide synchrotron light with certain polarization degree, it is at the cost of photon flux. According to the theory of SR, the electric field quantity that the bending magnet radiates could be decomposed into two components such as: $E_x$ and $E_y$, parallel to and vertical to the electron orbit. The phase difference between two components is 90°. Moreover, the horizontal component is larger than the vertical component. Their

relative amplitudes:

$$\begin{pmatrix} E_x \\ E_y \end{pmatrix} = \begin{pmatrix} \sqrt{1+(\gamma\psi)^2} \cdot K_{2/3}(\eta) \\ i\gamma\psi \cdot K_{1/3}(\eta) \end{pmatrix} \quad (1)$$

in which, $K_{1/3}$ and $K_{2/3}$ refer to the modified Bessel functions; y refers to the ratio of photon energy to the critical energy; $\gamma$ refers to the ratio of the electron energy to its rest mass energy; $\eta = (y/2)[1+(\gamma \cdot y)^2]^{3/2}$, and $\Psi$ refers to the vertical acceptance angle. The definition of $r$ is the ratio of the minor to major axis of the polarization ellipse, namely $r = E_y/iE_x$, and then the circular polarization degree could be gained: $P_c = 2r/(1+r^2)$ [1].

With the increase of $\Psi$, $P_c$ increases rapidly and tends to 1. By combining the result of Fig. 3, it indicates that there is a direct corresponding relationship between the value of $I_{up}/I_{down}$ and SR polarization. Thus the polarization variation could be directly estimated through measuring the values of $I_{up}$ and $I_{down}$.

It is noteworthy that, the change of vertical position for the track is often accompanied with the change in beam direction, as the convex rail system employed to adjust the beam position is composed of three magnets. Similar to the beam position, the beam direction also causes the change in incident light intensity and polarization, so the actual measurement result is the combined action of such two factors. With the continuous movement of the beam position as shown in Fig. 2, the SR intensity satisfies perfectly the Gaussian distribution. It seems that, compared with the beam position, the change of beam direction has little influence on light intensity, for it does not cause a disturbance to the light intensity with perfect Gaussian distribution. In terms of the polarization degree, the same beam position with different beam directions would change the value of $I_{up}/I_{down}$, and it also would change the circular polarization degree of synchrotron light entering the vacuum chamber. Fig. 4 displays several different situations of the light source and the chopper. Herein, $h$ refers to the electron offset distance from the orbit plane; $L$ refers to the horizon distance between the light source point and the chopper; $H$ represents the distance of the adjacent aperture at chopper; the dash dot line and dash line are the orbit plane and the beam direction respectively. Both ψ1 and ψ2 are the vertical acceptance angle of the beam passing through the upper and the lower apertures respectively. Fig. 4(a) is the most ideal case, there is no offset from the orbit plane and the beam direction is parallel to the orbit plane, i.e., ψ1 ψ2; while there is only orbital offset or beam direction change in Fig. 4(b) and Fig. 4 (c) respectively. In Fig. 4 (d), there exist orbital offset and beam direction change concurrently, representing the usual case.

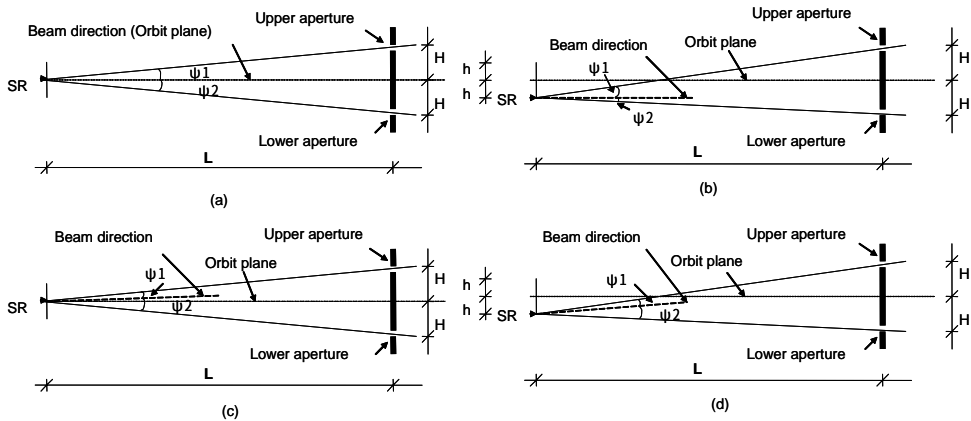

**Fig. 4.** The different situations of the light source and the chopper. (a) is the most ideal case; (b) refers to the case only with orbit offset; (c) refers to the case only with the beam direction change; (d) includes both orbit offset and beam direction change, representing the most usual case.

Either orbital offset or beam direction change can cause the change of vertical acceptance angle ψ, and further influence the polarization degree of the beam passing through the chopper. According to mathematics knowledge, considering h<<L, the sum of ψ1 and ψ2 basically remains unchanged. Fig. 5 shows the radiation intensity distribution as a function of ψ, once the ψ1 is determined, the polarized degree of beam entering the upper aperture is fixed; combined with the sum of ψ1 and ψ2 keeping a constant, the ratio of light flux passing through the upper and lower apertures is also fixed. Thus, whether the changes of $I_{up}/I_{down}$ come from beam position or beam direction, the polarization degree always maintains a corresponding relationship with the values of $I_{up}/I_{down}$. As the SR has the properties of decaying over time, using the above ratio as an indicator can effectively remove the influence caused by the SR decay. Compared with the beam position monitor, the values of $I_{up}/I_{down}$ includes the contributions from the orbital offset and the beam direction change, so it can be used to evaluate the SR polarization degree, providing the route for us to quantitatively study the circular polarization of SR entering the vacuum chamber with the value of $I_{up}/I_{down}$.

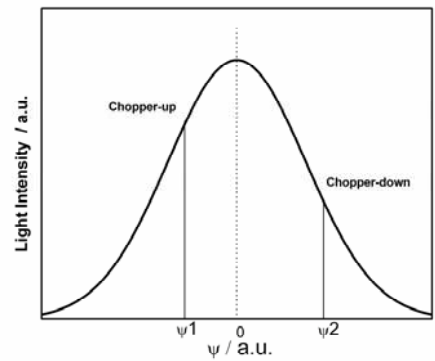

**Fig. 5.** Light intensity distribution of SR as a function of ψ.

# 4   Conclusions

After adjusting the photon beam position, it is found that the SR intensity variation at the light intensity monitor shows very good Gaussian distribution. Furthermore, the monotonous relationship between the light intensity ratio of upper and lower apertures and the vertical position of the photon beam is observed. This phenomenon indicates that we can estimate the relative changes of the polarization degree by measuring this intensity ratio, which provides an effective way for in-situ study of the SR polarization for the XMCD experiment.

**Expectation:** Next, we plan to measure the magnetic circular dichroism effect of the standard ferromagnetic sample, and calculate the polarization degree of incidence circular polarized light. Based on this, we will establish the correlation between the value of $I_{up}/I_{down}$ and the circular polarization degree of the incident light.

**Acknowledgments**

*We wish to thank all members of the SXMCD station at NSRL for their assistance during the experiments. Thanks also got to Professor Sun Baogen for providing constructive comments.*